\documentclass[%
 reprint,
 amsmath,amssymb,
 aps,
 prx,
]{revtex4-2}

\usepackage{graphicx}
\usepackage{dcolumn}
\usepackage{bm}

\usepackage{xcolor}
\usepackage{mathtools}
\usepackage{nicefrac}
\usepackage{subcaption}

\newcommand{\pc}{p_{\rm max}}
\newcommand{\dG}{\Delta G}
\newcommand{\com}[1]{}

\newcommand{\Tom}[1]{\color{blue}#1\color{black}}

\newcommand{\comment}[1]{ }

\begin{document}

\preprint{APS/123-QED}

\title{
Thermodynamic limits on general far-from-equilibrium molecular templating networks
}

\author{Benjamin Qureshi}
 
\author{Thomas E. Ouldridge}%
\email{t.ouldridge@imperial.ac.uk}

\affiliation{Department of Bioengineering and Centre for Synthetic Biology, Imperial College London, London SW7 2AZ, United Kingdom}

\author{Jenny M. Poulton}
\affiliation{Department of Physics and Astronomy, University of Sheffield, Sheffield S3 7RH, United Kingdom}

\date{\today}

\begin{abstract}
Cells produce RNA and proteins via molecular templating networks. We show that information transmission in such networks is bounded by functions of a simple thermodynamic property of the network, regardless of complexity. Surprisingly, putative systems operating at this bound do not have a high flux around the network. Instead, they have low entropy production, with each product in a ``pseudo-equilibrium'' determined by a single pathway. These pseudo-equilibrium limits constrain information transmission for the overall network, even if individual templates are arbitrarily specific.
\end{abstract}

\maketitle


Cellular mRNA and DNA molecules selectively catalyse the formation of many distinct protein and mRNA sequences from a small set of monomer building blocks~\cite{crick1970central,Ouldridge2017Fundamental}. Proteins and RNA are also both actively degraded~\cite{watson2014molecular}. The underlying processes are complex, with many pathways to assembly and disassembly and non-trivial motifs -- such as kinetic proofreading \cite{hopfield1974kinetic,ninio1975kinetic} -- within those pathways.  The net result is a distribution of protein and RNA sequences in the cell, biased towards template-specified targets; sharply-peaked, or {\it accurate}, distributions are essential for function. 

 The theory of such systems has focussed on  templating events in isolation~\cite{sahoo2021accuracy,song2020thermodynamic, li2019template,wong2018energy,bennett1979dissipation,banerjee2017elucidating,chiuchiu2019error,gaspard2014kinetics, poulton2019nonequilibrium,sartori2015thermodynamics,qureshi2023universal,gaspard2020template,juritz2022minimal,guntoro_interplay_2025}, rather than full networks. 
Refs.~\cite{Ouldridge2017Fundamental,bennett1982thermodynamics} hypothesised that such networks would require a minimal entropy production per output polymer, determined by the sequence information transmitted from template to products. Genthon {\it et al}.~\cite{genthon2024non} modelled a network with competing templating and spontaneous degradation pathways, treating both as a one-step process with an imposed kinetic selectivity. They observed a phase transition to a high accuracy regime with a large cyclic flux of templated production and spontaneous degradation, with a fuel turnover per product commensurate with the accuracy of the product ensemble, apparently confirming the predictions of \cite{Ouldridge2017Fundamental,bennett1982thermodynamics}.

Given the simplicity of the system in \cite{genthon2024non}, the thermodynamics of arbitrarily complex molecular templating networks is under-explored. Moreover, Refs.~\cite{Ouldridge2017Fundamental,bennett1982thermodynamics,genthon2024non}  suggest a paradox. Entropy production is related to the relative rates of time-reversed trajectories~\cite{seifert2012stochastic}, rather than the relative rates of distinct  processes. Although biochemistry may practically limit relative templating rates for matching and non-matching sequences, thermodynamics places no limit on this kinetic discrimination  {\it in principle} \cite{poulton2021edge}. How this fact is reconciled with a minimal cost of accurate templating networks is unclear.

In this letter, we first show that functions of a simple quantity -- $\dG$, the difference between the maximal and minimal free-energy changes along assembly pathways -- bound the accuracy of arbitrarily complex molecular templating networks. We then explore these bounds for a system in which $M$ possible products can be formed. For $M~\rightarrow~\infty$, it is  possible to maintain a steady-state ensemble with only a single product type, provided $\nicefrac{\Delta G}{\ln{M}} > k_{\rm B}T$. By contrast, a single product necessarily has zero weight within the ensemble for $\nicefrac{\Delta G}{\ln{M}} < k_{\rm B}T$. These results appear to confirm the hypotheses of Refs. \cite{bennett1982thermodynamics, Ouldridge2017Fundamental} and generalize Ref.~\cite{genthon2024non}'s result. Notably, however,  the bounds are more restrictive at finite $M$ and, for $M\rightarrow \infty$, one can surprisingly maintain an ensemble dominated by a vanishingly small fraction of the possible products for any $\Delta G \gg k_{\rm B}T$, even if $\Delta G \ll k_{\rm B}T\ln{M}$. 

\begin{figure}
    \includegraphics[scale=0.29]{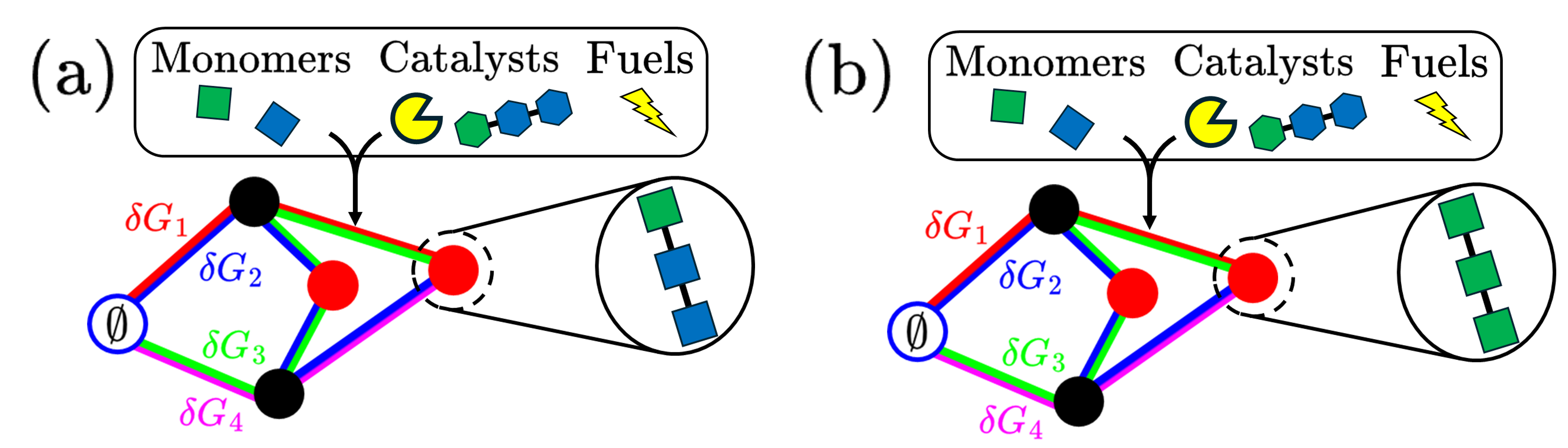}
    \caption{Modelling approach. Polymer products  are produced via catalysed addition of monomers, driven by fuel consumption. (a) Representation of the model as a graph, including a null complex $\emptyset$; intermediate species (black circles); and products (red circles). Each transition is reversible, and each distinct pathway from $\emptyset$ to the product has an associated free-energy change $\delta{G}_i$. In (b) we take an alternative product as the root of the graph; the network connecting it to the null state is topologically identical to (a). The equivalent pathways for each product have the same overall $\delta G_i$, but intermediate free energies and transition rates can vary.}
    \label{fig:schematic}
\end{figure}

Most significantly, however, systems that approach the $\dG$-dependent bounds do not have large, entropy-producing cyclic fluxes, as anticipated~\cite{Ouldridge2017Fundamental,bennett1982thermodynamics,genthon2024non} and observed in biology. Instead,  each product overwhelmingly couples to a single pathway, balancing forwards and backwards transitions and exhibiting a yield determined by that pathway's free-energy change.
These bound-saturating ``pseudo-equilibrium'' systems have negligible entropy production per assembly event, and  resolve the apparent paradox of the minimal cost of templating.

{\it Modelling framework.}
We consider a  broad class of networks, including that of Ref.~\cite{genthon2024non} as a special case. As shown in Fig.~\ref{fig:schematic}, we consider a set of monomer species ({\it e.g.} amino acids) that may assemble into $M$ possible products ({\it e.g.} protein sequences).  We assume that monomers are held at the same constant concentration (chemostatted)   by the environment. There are also catalysts ({\it e.g.} mRNA templates, ribosomoes and proteases) that facilitate the (dis)assembly of products via intermediate states, and fuel molecules that can drive reactions. Catalysts and fuels are also chemostatted, and we neglect reactions of more than one product and/or intermediate. 

We analyse these systems at the level of deterministic chemical reaction networks (CRNs)~\cite{polettini2014irreversible,rao2016nonequilibrium} obeying mass-action kinetics.
CRNs consist of a set of species, a set of complexes (collections of species), and a set of reactions in which a complex $\mathcal{X}$ is converted into a complex $\mathcal{Y}$ at a rate given by a rate constant $k_{\mathcal{X}\to \mathcal{Y}}$ multiplied by the product of concentrations of the species in $\mathcal{X}$. 

Under our assumptions, only the products and intermediates are treated explicitly as species, with  monomer, fuel and free catalyst concentrations acting as parameters that are absorbed into rate constants. 
Each complex then either contains a single species or is null ($\emptyset$); transitions from  $\emptyset$ correspond to the first association between monomers and catalysts.  The CRNs are therefore linear and can be drawn as a graph (Fig.~\ref{fig:schematic}\,(a)), with nodes as complexes, edges as reactions and weights as reaction rates. The graph is connected, since all complexes can be reached from $\emptyset$. This graph defines a set of self-avoiding walks (SAWs) from $\emptyset$ to each product; each SAW and its inverse define a ``pathway''. 
We assume that all products are connected to $\emptyset$ by topologically equivalent graphs (Fig.~\ref{fig:schematic}\,(b)). Heuristically, every pathway from $\emptyset$ to a given product has an equivalent for every other product that couples to the same catalysts and consumes the same fuel but incorporates different monomers, a reasonable assumption for fixed-length copolymers (see section I of \cite{our_appendix} for a more formal definition).

{\it Model thermodynamics}. For thermodynamic self-consistency, we require that forward ($\mathcal{X}\to \mathcal{Y}$) and reverse ($\mathcal{Y} \to \mathcal{X}$) transitions obey local detailed balance \cite{seifert2012stochastic}: $\delta \tilde{G}_{\mathcal{X}\to \mathcal{Y}} = -\ln\left({k_{\mathcal{X}\to \mathcal{Y}}}/{k_{\mathcal{Y}\to \mathcal{X}}}\right)$, where $k_BT \delta \tilde{G}_{\mathcal{X}\to \mathcal{Y}} $ is the standard free energy change of reaction \cite{ouldridge2018importance}.  Here, $\delta \tilde{G}_{\mathcal{X}\to \mathcal{Y}} =  \tilde{G}_{\mathcal{Y}}-\tilde{G}_{\mathcal{X}} +\sum_i \tilde{\mu_i} \delta N^{\mathcal{X}\to \mathcal{Y}}_i$, with $\tilde{G}_{\mathcal{X}}-\tilde{G}_{\mathcal{Y}}$  the free-energy change of the non-chemostatted species, and  $\tilde{\mu_i} \delta N^{\mathcal{X}\to \mathcal{Y}}_i$ the free-energy change due to the production of $\delta N^{\mathcal{X}\to \mathcal{Y}}_i$ molecules of type $i$, chemostatted at  chemical potential $k_BT \tilde{\mu}_i$. Each edge corresponds to a fixed $\delta N^{\mathcal{X}\to \mathcal{Y}}_i$, so  $\delta \tilde{G}_{\mathcal{X}\to \mathcal{Y}} $ is well defined. However, two SAWs $S$ and $S^\prime$ connecting  $\emptyset$ and a given product can consume different amounts of fuel, allowing $\delta \tilde{G}_S \neq \delta \tilde{G}_{S^\prime}$, where  $\delta \tilde{G}_S =\sum_{e \in S}  \delta \tilde{G}_{e}$ (Fig.~\ref{fig:schematic}\,(a)). The framework thus incorporates arbitrarily complex non-equilibrium networks including, for example, kinetic proofreading cycles.

Templating is catalytic, so product stability is template independent \cite{poulton2019nonequilibrium,Ouldridge2017Fundamental}. The simplest consistent approach  is to  assume all products have the same standard free energy  $\tilde{G_0}$, regardless of the templates~\cite{Ouldridge2017Fundamental,poulton2019nonequilibrium,poulton2021edge,juritz2022minimal,genthon2024non}. The total free energy change along a SAW is then $\delta \tilde{G}_S =\tilde{G_0} + \sum_{e \in S} \sum_i \tilde{\mu_i} \delta N^e_i$. Since all products are connected to $\emptyset$ by topologically equivalent pathways involving the same fuel consumption, the set of  free-energy changes associated with the formation of each product is also equivalent  (Fig.~\ref{fig:schematic} and section I of \cite{our_appendix}). By contrast, we allow for arbitrary intermediate free energies, and transition rates are unconstrained except by $\delta \tilde{G}_{\mathcal{X}\to \mathcal{Y}} = -\ln\left({k_{\mathcal{X}\to \mathcal{Y}}}/{k_{\mathcal{Y}\to \mathcal{X}}}\right)$. Certain products can then form faster via selective templating.




 We first show that, under these assumptions, properties of the steady-state distribution of products are bounded by functions of $\Delta \tilde{G}$, the difference between the maximum and minimum $\delta \tilde{G}_S$. We will then explore the nature of these bounds.

{\it Bounds on the product ensemble.}  
We consider the steady-state concentration of products (the ``product ensemble''); our results also apply to the expected steady-state concentrations of a stochastic realisation of the CRN~\cite{anderson2010product}. For linear, connected  CRNs with a null complex, the steady-state concentration of any species is bound by the free-energy changes along the SAWs connecting that product state to $\emptyset$  \cite{maes2013heat,saez2019linear,nam2022linear,cetiner2022reformulating}. 
Defining $\delta \tilde{G}^{Z_i}_L=\max\limits_{S} \delta \tilde{G}^{Z_i}_S$ and $\delta \tilde{G}^{Z_i}_U=\min\limits_{S} \delta \tilde{G}^{Z_i}_S$, where the optimization is performed over all SAWs that lead to $Z_i$ from $\emptyset$, then
$
    e^{-\delta \tilde{G}^{Z_i}_L}\leq c_{Z_i} \leq e^{-\delta \tilde{G}^{Z_i}_U}
$
(see section II of \cite{our_appendix}).
For our networks, all products are connected to $\emptyset$ by pathways with the same set of free-energy changes. All products therefore have the same upper and lower bounds ($e^{-\delta \tilde{G}_U},e^{-\delta \tilde{G}_L}$), a crucial fact placing fundamental constraints on the ensemble.

To see how these constraints manifest, we define the product distribution $\mathbb{P}(Z_i)$ for products $Z_i$,  $i=1,\cdots M$:
\begin{equation}
    \mathbb{P}(Z_i)=p_i=\nicefrac{c_{Z_i}}{\sum\limits_{j=1}^Mc_{Z_j}} = \nicefrac{c_{Z_i}}{c_T}.
    \label{eq:dist}
\end{equation}
Here, $c_{Z_i}$ is the concentration of product $Z_i$ and $c_T=\sum_{j=1}^Mc_{Z_j}$ is the total concentration. We consider two metrics for the deviation from a uniform equilibrium ensemble. First, the single product  specificity  $\pc=\max\limits_{i}p_i$, which gives the degree to which the a single template-specified product dominates the ensemble. Second, the (Shannon) entropy~\cite{shannon1948mathematical} of the distribution,
\begin{equation}
    H[p_i]=-\sum\limits_{i=1}^Mp_i\ln{p_i}=\ln{c_T}-\frac{1}{c_T}\sum\limits_{i=1}^Mc_{Z_i}\ln{c_{Z_i}}.
\end{equation}
$\ln M - H[p_i]$ 
is the channel capacity of the network, if it is treated as an information channel from templates to products  (see section IV of \cite{our_appendix}). We describe optimizing for single product specificity or entropy as {\it specificity maximisation} or {\it entropy minimization}, respectively.

\begin{figure*}
    \centering
        \includegraphics[scale=0.27]{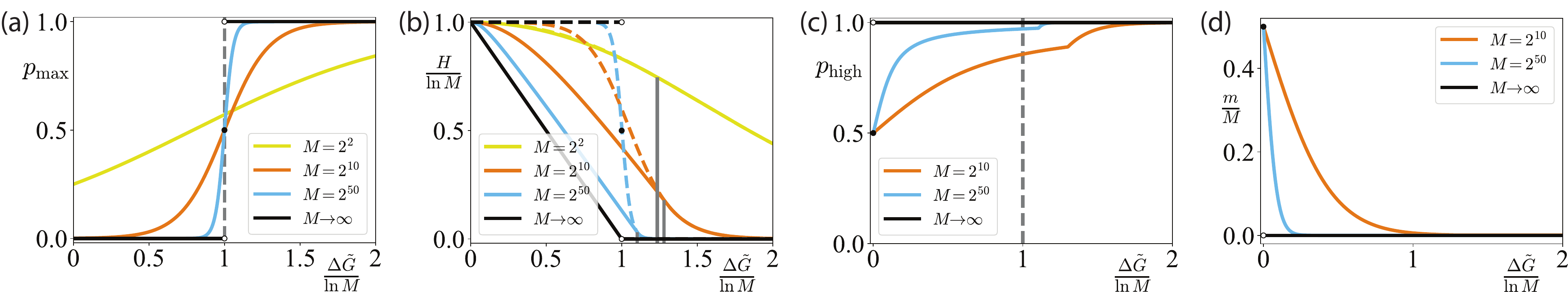}
    \caption{Behaviour of single-product specificity and entropy bounds. (a) The bound on single-product specificity $p_{\rm max}$ as a function of $\Delta \tilde{G}/\ln{M}$, for various $M$. (b) The bound on entropy $H_{\rm min}$ (solid lines) and the entropy of the distribution of maximal single-product specificity (dotted lines) as a function of $\Delta \tilde{G}$ for various $M$. We scale both quantities by $\ln{M}$. The dashed line is discontinuous at $\Delta \tilde{G} = \ln M$ for $M \rightarrow \infty$. The vertical (gray) lines show $\Delta \tilde{G} = \ln M+\ln\ln M$, below which the curves do not overlap. (c) The fraction of products at the high concentration in an entropy-minimized ensemble, $m_{\rm min}/M$, as a function of $\Delta \tilde{G}/\ln{M}$, for various $M$. (d) The probability of selecting a product at the high concentration in the entropy-minimized ensemble, $p_{\rm high}$, as a function of $\Delta \tilde{G}/\ln{M}$, for various $M$. A small fraction of products can dominate the ensemble, even below  $\Delta \tilde{G}/\ln{M}=1$.}
    \label{fig:graphs}
\end{figure*}

Given \mbox{$e^{-\delta \tilde{G}_L}\leq c_{Z_i}~\leq e^{-\delta \tilde{G}_U}$}, specificity maximisation at fixed $\Delta \tilde{G} = \delta \tilde{G}_L-\delta \tilde{G}_U$ is saturated by an ensemble with one species at  $\pc$, and all others at $p_{\rm low}$, with
\begin{eqnarray}
    \pc=\left(1+(M-1)e^{-\Delta \tilde{G}}\right)^{-1}, 
    p_{\rm low}=e^{-\Delta \tilde{G}}\pc.
\label{eq:pmax}
\end{eqnarray}
This $\pc(\Delta \tilde{G})$ thus defines a bound. {Entropy minimisation} yields $H[p_i] \geq H(m)$, with
\begin{equation}
    H(m) =\frac{(M-m)\Delta \tilde{G} e^{-\Delta \tilde{G}}}{m+(M-m)e^{-\Delta \tilde{G}}}+\ln\left(m+(M-m)e^{-\Delta \tilde{G}}\right).
    \label{eq:Hmin}
\end{equation}
$H(m)$ corresponds to $m$ species at the upper bound, $c_U=\exp(-\delta \tilde{G}_U)$, and $M-m$ species at the lower bound, $c_L=\exp(-\delta \tilde{G}_L)$. Minimizing Eq.~\ref{eq:Hmin} with respect to integer $m$ yields the lower bound on entropy, $H_{\rm min}(\Delta \tilde{G})$. $H_{\rm min}$ and the minimizing $m=m_{\rm min}$, and a proof of Eq.~\ref{eq:Hmin}, are given in section III of \cite{our_appendix}. For illustrative purposes,  we use  the approximation in Eq.~S17, which treats $m_{\rm min}$ as continuous when $m_{\rm min}>1$ (a slightly weaker bound).


{\it  Specificity maximisation and entropy minimisation.}
In Fig.~\ref{fig:graphs}\,(a), we plot the bound on $p_{\rm max}$ against $\Delta \tilde{G}/\ln M$ for various $M$ ($\ln M$ is proportional to the length of a copolymer). $p_{\rm max}$ undergoes a transition from $M^{-1}$ to 1 as $\Delta \tilde{G}/\ln M$ is increased, centred on $\Delta \tilde{G}/\ln M=1$. As $M\rightarrow \infty$, the sigmoid becomes a phase transition \cite{genthon2024non}.

Surprisingly, specificity maximisation and entropy minimisation are not equivalent. In Fig.~\ref{fig:graphs}\,(b), we plot the bound on $H_{\rm min}$ and the entropy of the maximum specificity distribution  against $\Delta \tilde{G}/\ln M$. The two curves overlap for  $\Delta \tilde{G} > \ln M +\ln\ln M + \mathcal{O}\left(\frac{\ln\ln M}{\ln M}\right)$, when $m=1$ minimises Eq.~\ref{eq:Hmin} and the lowest entropy state has a single product at a concentration $c_U=e^{-\delta \tilde{G}_U}$, and all others at $c_L=e^{-\delta \tilde{G}_L}$. For smaller $\Delta \tilde{G}$, the two differ drastically. $H_{\rm min}$  is obtained for distributions with $m_{\rm min}>1$ products at $c_U$. This unexpected behaviour arises because having $m>1$ products at $e^{-\delta \tilde{G}_U}$ increases $c_T$ and thus suppresses the probabilities of other species.

Fig.~\ref{fig:graphs}\,(c) shows the fraction $m_{\rm min}/M$ of products at $c_U$ in the entropy-minimizing distribution, and Fig.~\ref{fig:graphs}\,(d) the total probability $p_{\rm high}$ of selecting a high concentration product from that ensemble.
Although $m_{\rm min}$ increases and $p_{\rm high}$ decreases as $\Delta \tilde{G}\rightarrow 0$, for large $M$, $m_{\rm min}/M$  tends to zero while maintaining $p_{\rm high}>0.5$ even for $\Delta \tilde{G}\ll\ln{M}$, well inside the region where single product specificity is impossible.  As $M\rightarrow \infty$ at fixed  $\nicefrac{\Delta \tilde{G}}{\ln{M}}$, a vanishingly small proportion of the total number of possible products ($m_{\rm min}/M\rightarrow 0)$ can dominate the ensemble ($p_{\rm high} \rightarrow 1$) even when $\Delta \tilde{G} \ll \ln M$. 

{\it Physical meaning of the bounds.} 
The phase transition in $p_{\rm max}$ at $\nicefrac{\Delta \tilde{G}}{\ln{M}}=1$ for $M \rightarrow \infty$ apparently generalises the observations of Ref.~\cite{genthon2024non} to arbitrarily complex networks and supports the hypotheses of \cite{Ouldridge2017Fundamental,bennett1982thermodynamics}, suggesting a minimal cost of accurately copying a single template.
Notably, however, Fig.~\ref{fig:graphs}\,(c,d) are hard to interpret in the context of a cost of accuracy, as they  suggest that a relatively small set of templates can be copied precisely at negligible cost. Simultaneously, significantly larger values of $\nicefrac{\Delta \tilde{G}}{\ln{M}}$ are required to give perfect accuracy for finite $M$ (see Fig.~\ref{fig:graphs}\,(a)). Consider the implications for the charging of a single tRNA with an amino acid, an example of templated dimerization. There are approximately $M=400$ combinations of codon and amino acid, having grouped redundant codons. For a single charged tRNA to dominate the ensemble with an error rate of $10^{-5}$, one would need $\Delta \tilde{G}$ of around $17k_BT$ or 1 ATP at $37^\circ C$, substantially higher than $kT\ln M = k_BT \ln 400 \approx 6k_BT$, the value implied by the large $M$ limit.

\begin{figure}
    \includegraphics[scale=0.25]{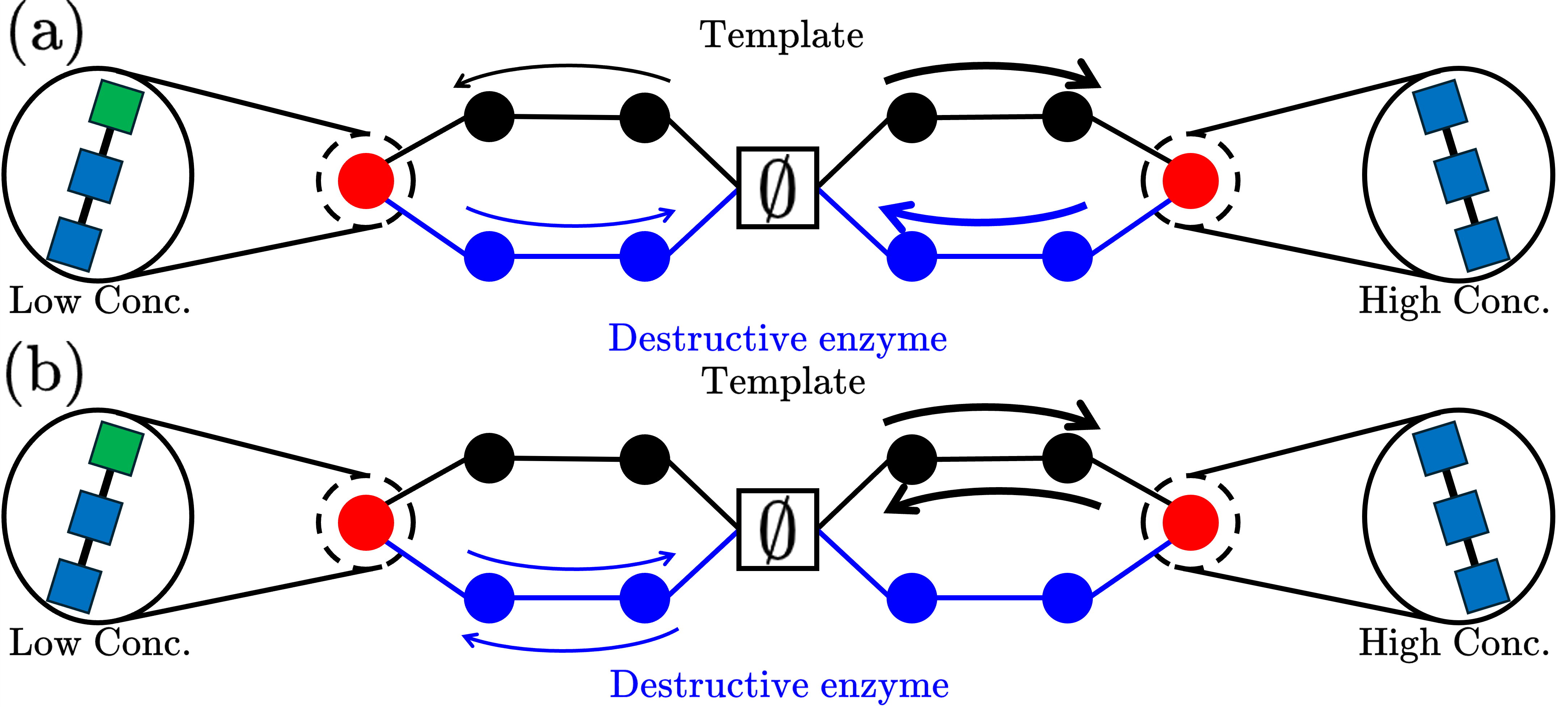}
    \caption{Comparison of (a) a steady state with high cycle flux and (b) pseudo-equilibrium, showing the dominant production and destruction trajectories only.}
    \label{fig:schematic2}
\end{figure}

Most importantly, a putative system that approaches the bounds on $\pc$ and $H[p_i]$ behaves unexpectedly (Fig.~\ref{fig:schematic2}). 
 $c_{Z_X}=\exp(-\delta \tilde{G}_U)$ can only be reached if $Z_X$ is  produced and degraded solely by the pathway with free-energy change  $\delta \tilde{G}_U$; similarly,  $c_{Z_Y}=\exp(-\delta \tilde{G}_L)$ requires $Z_Y$ to be produced and degraded solely by the pathway with free-energy change  $\delta \tilde{G}_L$.  $Z_X$ and $Z_Y$ are then in ``pseudo-equilibria'', with their yields equal to the equilibrium yield of the relevant pathway.


These pseudo-equilibria, whether or not they are practically achievable, limit all possible non-equilibrium templating systems for a given $\Delta \tilde{G}$. Indeed, any non-zero net flux necessarily {\it reduces} accuracy relative to this limit. Consider two products $X$ and $Y$, and assume $c_{X}>c_{Y}$. Let $\Gamma_X(\delta \tilde{G})$ be the observed ratio of production to degradation trajectories of $X$ along a pathway with free-energy change $ \delta \tilde{G}$; deviations of $\Gamma$ from unity imply a net flux. At steady state, thermodynamic self-consistency implies $ \Gamma_X(\delta \tilde{G}) =\exp(-\delta \tilde{G})/c_{X}$. Thus, 
\begin{eqnarray}
    \ln\left({c_{X}}/{c_{Y}}\right) = \Delta \tilde{G} +  \ln\left({\Gamma_Y(\delta \tilde{G}_L)}/{\Gamma_X(\delta \tilde{G}_U)}\right).
\end{eqnarray}
Since $\delta \tilde{G}_U$ corresponds to the most favourable production pathway, $\Gamma_X(\delta \tilde{G}_U)\geq 1$. Conversely, $\Gamma_Y(\delta \tilde{G}_L)\leq 1$. We see that a systematic flux to produce $X$ along the most favourable production pathway, or degrade $Y$ along the least favourable production pathway, necessarily reduces the concentration ratio beneath that required to reach the bounds ($c_{X} = c_{Y} \exp(\Delta \tilde{G})$ if $c_{X}> c_{Y}$). 

Equally surprisingly, while operating at this pseudo-equilibrium limit requires a certain $\Delta \tilde{G}$ for accuracy, this ``cost'' is not an entropy production per product made, as it would be for a system with a high cyclic flux. If typical production and degradation events are truly time-reversed processes, the system will have zero net fuel consumption; all fuel consumed in production would be regenerated during destruction. Entropy generated per production or degradation event is thus zero in steady state -- at variance with \cite{Ouldridge2017Fundamental,bennett1982thermodynamics,genthon2024non}.

Fundamentally, templating's limiting ``cost'' is not needed to ensure selective catalysis -- the ratio of catalytic rates is unconstrained by the second law. Rather, $\Delta \tilde{G}$-dependent bounds arise because steady-state yields cannot vary by more than the pseudo-equilibria arising from maximally different pathways. This observation explains why accurate templating systems have a minimal cost despite apparently only needing to distinguish between processes that are not time-reverse pairs.

Pseudo-equilibrium limits only apply in steady state, and  no lower bound on $\Delta \tilde{G}$ is required for accuracy outside of steady state. Consider a system with two products:
$
    \emptyset\xrightleftharpoons[k_{X}]{k_{X}}X,\;\emptyset\xrightleftharpoons[k_{Y}]{k_{Y}}Y.
$
The steady state ensemble is unbiased, $p_{X}=p_{Y}=1/2$, since $\Delta \tilde{G}=0$. However, $c_{X}$ and $c_{Y}$ can be very different at short times. Starting with initial conditions $c_{X}(0) = c_{Y}(0)=0$, 
$
    \lim_{t\to0} p_X(t) = \frac{k_X}{k_X+k_Y},
$
which is only bounded by 0,1. A catalyst that prefers $X$ can temporarily achieve arbitrary specificity in the product ensemble even though $\Delta \tilde{G}=0$, as demonstrated in Ref.~\cite{Cabello2025information}.

We have derived thermodynamic bounds on the accuracy of arbitrarily complex catalytic molecular templating networks, and shown that systems at these bounds operate in a pseudo-equilibrium fashion with negligible cyclic flux. Our results raise the question of why cells do not operate in this way, which is accurate and low cost. First, we have derived bounds, not constructed realistic systems that approach them. In Section V of \cite{our_appendix} we give an explicit kinetic model of a templating network that saturates the bounds as rate constants are tuned. However, the network is contrived, and it is hard to imagine a system in which, for example, production and degradation by an RNA polymerase are balanced. Moreover, not all networks can reach the bounding accuracy, even if rate constants in a network can be tuned arbitrarily. We provide an example in section VI of \cite{our_appendix}. 

There are also  disadvantages of operating in the pseudo-equilibrium limit. First, outputs are binary:  the steady state concentration doesn't depend on the precise template concentration. Biological systems likely benefit from analog control. Second, fast switching between high and low states is more challenging when production and degradation aren't controlled by different catalysts. Synthetic information-processing systems, however may have different requirements from nature: a binary output is often preferable, products may be much simpler than long protein and RNA sequences, and fast switching may not be important. Synthetic pseduo-equilibrium templating networks may, therefore, hold promise. 

We have made a number of assumptions in our analysis. Although we consider a very general class of linear networks, it is unclear how these results transfer to the non-linear setting. The most likely violation of linearity is that templates and catalysts become depleted. In this case, however, pathways of well-defined $\delta \tilde{G}$ still exist, which may allow similar limits to be derived~\cite{liang2024thermodynamic}. We have also not considered the yield of partially formed  products, which will be relevant if templating is not highly processive, or if template-bound intermediates have downstream functionality.
Finally, we have assumed product equivalency. If $\delta \tilde{G}_U$ and $\delta \tilde{G}_L$ are product-specific, much of our analysis would still apply, with limiting accuracy obtained when  products are in pseudo-equilibrium at either their maximum ($e^{-\delta \tilde{G}_U}$) or minimum ($e^{-\delta \tilde{G}_L}$) concentrations. It will be easier to create distributions that favour stable products, and harder in other cases. Given that transcription and translation produce functional, rather than stable, products, it is unclear whether this asymmetry would be beneficial. 

{\it Acknowledgements.} We thank Pieter Rein ten Wolde for his  comments on the manuscript. This work is part of a project that has received funding from the European Research Council (ERC) under the European Union’s Horizon 2020 research and innovation program (Grant agreement No. 851910). T.E.O. was supported by a Royal Society University Research Fellowship.

{\it Data Availability Statement.}
Code supporting the findings of this study is openly available at \mbox{\url{https://zenodo.org/records/15554491}}.

\begin{appendix}
\renewcommand{\thesection}{S\Roman{section}} 
\renewcommand{\thefigure}{S\arabic{figure}} 
\renewcommand{\thesubsection}{\thesection.\Alph{subsection}} 
\renewcommand{\theequation}{S\arabic{equation}} 

\section{Definition of topological equivalence}
\label{App:topologicalequivalence}
Let $\mathcal{G}$ be the vertex-labelled graph describing the linear CRN containing a vertex corresponding to the null complex $\emptyset$ and a set of vertices corresponding to products, ignoring the edge-weights. Then for any pair of products, $X$ and $Y$, consider a relabelling of the nodes such that the nodes for $X$ and $Y$ swap labels, $\emptyset$'s label is not changed, but other node labels may be permuted. Suppose there exists such a relabelling, transforming the graph to a new vertex-labelled graph $\mathcal{G}'$ such that $\mathcal{G}'$ is isomorphic as a labelled graph to $\mathcal{G}$. Then the set of SAWs from $\emptyset$ to $X$ will be topologically equivalent to those SAWs from $\emptyset$ to $Y$. Further, a given SAW from $\emptyset$ to $X$ will be topologically equivalent to a specific SAW from $\emptyset$ to $Y$. If these equivalent SAWs also incur the same free-energy change, then the model falls into the framework of models we discuss in the paper.

\section{Proof of boundedness of steady-state concentrations}
\label{App:Proofs1}
\subsection{The steady state concentration written as sum over spanning trees}
\label{App:SpanningTree}
We first give a useful result for the concentration of a species in terms of the graph of the linear CRN. We note that a similar result can be seen in {\it e.g.} \cite{nam2022linear}. However, we prove this result here with particular reference to the effect of a null species.

Consider a linear connected CRN under mass action kinetics. Let $X_i$ be the chemical species with concentration $c_i$. Assume the CRN contains some reactions of the form $\emptyset\rightleftharpoons X$. Without loss of generality, we assume that there only exists at most one reaction of the form $X_i \to X_j$ (for the cases where there are multiple such reactions, replace their reaction rate with the sum over all reaction rates of reactions of that form). We may cast the steady state equation for the vector of steady state concentrations of the chemical species, $\mathbf{c}$, as the linear equation:
\begin{eqnarray}
    A\mathbf{c} + \mathbf{b}=\mathbf{0},
    \label{eq:LinearSS}
\end{eqnarray}
where $\mathbf{b}$ is a vector such that entry $b_i$ is the rate constant of the reaction $\emptyset\to X_i$, and $A$ is a matrix with off diagonal entries $A_{ij}$ equal to the rate constant of the reaction $X_j \to X_i$ and diagonal elements $A_{ii}$ equal to minus the sum of reaction rates of reactions $X_i\to X_j$ over all chemical species $X_j$ $j\neq i$ as well as $X_i \to \emptyset$. Note that the sum of column $i$ of matrix $A$ is equal to minus the rate constant of reaction $X_i\to \emptyset$. Thus, we may create the new matrix:
\begin{eqnarray}
   K=\left( \begin{array}{c|c}
        A&\mathbf{b}\\\\
        \hline\\
        \mathbf{d}^T&-\sum\limits_ib_i
    \end{array}
    \right),
\end{eqnarray}
where $\mathbf{d}$ is a vector such that entry $d_i$ is equal to the rate constant of reaction $X_i\to\emptyset$. The columns of the matrix, $K$, now sum to zero and can be recognised as the Laplacian matrix of a certain graph. Represent $K$ as a graph with nodes corresponding to chemical species $X_i$ and an additional node corresponding to $\emptyset$, and edges $e$ corresponding to reactions between species with weights equal to their rate constant $k(e)$. Then, by the matrix tree theorem~\cite{tutte1948dissection}, the determinant (up to a sign) of the sub matrix formed by deleting row and column $i$ from matrix $K$ is given by the sum over the set of spanning trees rooted at node $i$, $\mathcal{T}(X_i)$, 
\begin{equation}
    \det(K/i)=\sum\limits_{T\in\mathcal{T}(X_i)}\prod\limits_{e\in T}k(e),
\end{equation}
where $(K/i)$ represents matrix $K$ with row and column $i$ deleted. In particular, 
\begin{equation}
    \det(A) = \sum\limits_{T\in\mathcal{T}(\emptyset)}\prod\limits_{e\in T}k(e).
\end{equation}
Thus, by Cramers rule~\cite{Robinson1970Cramers}, and the sign of the determinant under swapping of rows, 
\begin{equation}
    c_X=\frac{\sum\limits_{T\in\mathcal{T}(X)}\prod\limits_{e\in T}k(e)}{\sum\limits_{T\in\mathcal{T}(\emptyset)}\prod\limits_{e\in T}k(e)},
    \label{eq:SpanTrees}
\end{equation}
gives the solution to eq.~\ref{eq:LinearSS}.

\subsection{Bounding steady-state concentrations}
We note that similar proofs exist in the literature \cite{maes2013heat,saez2019linear,nam2022linear,cetiner2022reformulating}, but we have included the proof here for completeness and to match our specific conventions. Further, in the cases considered in this paper, the fact that all products are connected to the null state means that our result bounds the absolute, rather than a relative, concentrations.

Consider a linear connected CRN with chemical species $X_i$ and some reactions of the form $\emptyset\rightleftharpoons X_i$. The concentration of species $X_i$ may be written as in eq~\ref{eq:SpanTrees}. The numerator and denominator in this fraction are sums over spanning trees rooted at a given node. A sum over spanning trees rooted at node $X$ may be factored into a sum over self avoiding walks (SAWs) from some arbitrary other node to node $X$. Concretely, letting $Y$ be the other, arbitrary node, and $S(Y\to X)$ be the set of SAWs from $Y$ to $X$,
\begin{equation}
    \sum\limits_{T\in\mathcal{T}(X)}\prod\limits_{e\in T}k(e) = \sum_{S\in S(Y\to X)}A(S)\prod_{e\in S}k(e),
\end{equation}
where $A(S)$ is a factor that, crucially, is the same for the equivalent (reversed) SAW in $S(X\to Y)$, in which all edges are reversed compared to $S(Y\to X)$. That is to say, if we now wish to find the sum over spanning trees rooted at $Y$, we may choose $X$ as the arbitrary other state and find:
\begin{equation}
    \sum\limits_{T\in\mathcal{T}(Y)}\prod\limits_{e\in T}k(e) = \sum_{S\in S(Y\to X)}A(S)\prod_{e\in S}k(\Bar{e}),
\end{equation}
where $\Bar{e}$ is the reverse of edge $e$. For the linear CRNs, utilising
\begin{equation}
    \delta \tilde{G}_S = -\ln\left(\prod\limits_{e\in S}\frac{k(e)}{k(\Bar{e})}\right),
    \label{eq:SAWdG}
\end{equation}
which follows from applying local detailed balance to each step of the SAW \cite{ouldridge2018importance}, we may write:
\begin{eqnarray}
    c_X &=& \frac{\sum\limits_{S\in S(\emptyset\to X)}A(S)\prod\limits_{e\in S}k(e)}{\sum\limits_{S\in S(\emptyset\to X)}A(S)\prod\limits_{e\in S}k(\Bar{e})}\nonumber\\
&=& \frac{\sum\limits_{S\in S(\emptyset\to X)}A(S)\left[\prod\limits_{e\in S}k(\Bar{e})\right]e^{-\delta \tilde{G}_S}}{\sum\limits_{S\in S(\emptyset\to X)}A(S)\prod\limits_{e\in S}k(\bar{e})}.
\end{eqnarray}
Hence,
\begin{equation}
    c_X\in\left[e^{-\max\limits_{S\in S(\emptyset\to X)}(\delta G_S)}, e^{-\min\limits_{S\in S(\emptyset\to X)}(-\delta G_S)}\right],
\end{equation}
as required.

\section{Proof of the boundedness of steady-state distribution entropy}
\label{App:Proofs2}
We have a set of $M$ concentrations $\{c_1,\dots c_M\}$. Denote the total concentration $c_T=\sum\limits_{i=1}^Mc_i$.
Suppose that each concentration is bounded by the same upper and lower bounds, $c_i\in[c_L,c_U]$. We now propose that the distribution of concentrations that minimises the Shannon entropy, $H([c_i])$, is that with $m_{\rm min}$ of the species at concentration $c_U$ and $M-m_{\rm min}$ at concentration $c_L$, where $m_{\rm min}$ is either
\begin{eqnarray}
    &&\left\lceil\frac{\frac{c_L}{c_U} \left[-\ln\left(\frac{c_L}{c_U}\right)-\left(1-\frac{c_L}{c_U}\right)\right]}{\left(1-\frac{c_L}{c_U}\right)^2}M \right\rceil \;\text{or}    \nonumber \\
    &&\left\lceil\frac{\frac{c_L}{c_U} \left[-\ln\left(\frac{c_L}{c_U}\right)-\left(1-\frac{c_L}{c_U}\right)\right]}{\left(1-\frac{c_L}{c_U}\right)^2}M \right\rceil-1.
\label{Eq:mInteger}
\end{eqnarray}

To prove this claim, let us calculate the derivative of $H=H([p_i])$ with respect to a concentration $c_\alpha$, holding all other concentrations fixed and remembering that $c_T$ is linear in $c_\alpha$,
\begin{equation}
    \frac{\partial H}{\partial c_\alpha}=\frac{1}{c_T}\left(-\ln\left(\frac{c_\alpha}{c_T}\right)-H\right).
\end{equation}
This $H$ has a local maximum or minimum at $-\ln \left(\frac{c_\alpha}{c_T}\right)=H$.
To proceed, we require the second derivative,
\begin{equation}
    \frac{\partial^2H}{\partial c_\alpha^2}=-\frac{1}{c_T^2}\left(-\ln\left(\frac{c_\alpha}{c_T}\right)-H\right)-\frac{1}{c_T}\frac{\partial H}{\partial c_\alpha}-\frac{1}{c_T}\frac{1}{c_\alpha}\left(1-\frac{c_\alpha}{c_T}\right).
\end{equation}
Evaluating the second derivative at $-\ln\left(\frac{c_\alpha}{c_T}\right)=H$, the first two terms are zero and the third is necessarily negative for non-zero $c_T$. Thus
\begin{equation}
    \left. \frac{\partial^2H}{\partial c_\alpha^2} \right \vert_{-\ln\left(\frac{c_\alpha}{c_T}\right)=H}<0.
\end{equation}
And so, for any distribution, we can decrease $H$ by increasing the concentrations of any species $i$ for which $-\ln\left(\frac{c_i}{c_T}\right)<H$ and decreasing the concentrations for any species whose concentration has $-\ln\left(\frac{c_i}{c_T}\right)>H$. For species for which $-\ln\left(\frac{c_i}{c_T}\right)=H$, changing the concentration in either direction will decrease $H$. Consequently, minimising the entropy of the distribution necessarily requires all species to be at one bound or the other: we need $m$ species at concentration $c_U$ and $M-m$ at concentration $c_L$. Hence, we have transformed the problem into a one dimensional one of minimising $H$ as a function of $m$. We can write this entropy as
\begin{equation}
    H(m)=-\frac{(M-m)\frac{c_L}{c_U}\ln \left(\frac{c_L}{c_U}\right)}{(M-m)\frac{c_L}{c_U}+m}+\ln\left((M-m)\frac{c_L}{c_U}+m\right).
\end{equation}
Taking the derivative with respect to $m$ and setting it to zero tells us that 
\begin{equation}
    m_{\rm min}=\frac{\frac{c_L}{c_U} \left[-\ln\left(\frac{c_L}{c_U}\right)-\left(1-\frac{c_L}{c_U}\right)\right]}{\left(1-\frac{c_L}{c_U}\right)^2}M
\end{equation}
gives a turning point for $H(m)$, which is clearly a minimum since $H(0)=H(M)=\ln{M}$ is maximal entropy. Since $H(m)$ has only a single turning point as a function of $m$, the integer that minimises $H(m)$ will be either the floor or ceiling of the above expression, and $H_{\rm min}$ is given by eq. 4 of the main text and Eq.~\ref{Eq:mInteger} (remembering that $c_L/c_U = \exp(-\Delta \tilde{G})$).

A good approximation to the optimal value of $m$ is given by
\begin{equation}
    m^\prime_{\rm min}=\max\left(\frac{e^{-\Delta \tilde{G}} \left[\Delta \tilde{G}-\left(1-e^{-\Delta \tilde{G}}\right)\right]}{\left(1-e^{-\Delta \tilde{G}}\right)^2}M,1\right),
    \label{Eq:HighFrac}
\end{equation}
since for $m_{\rm min}>1$, the difference between using the integer value of $m_{\rm min}$ (eq.~\ref{Eq:mInteger}) and the continuous value $m^\prime_{\rm min}$ (eq.~\ref{Eq:HighFrac}) is small. In practice, using equation~\ref{Eq:HighFrac} produces a very slightly looser bound on $H[p_i]$. Further, using eq.~\ref{Eq:HighFrac}, for $m_{\rm min}>1$, we may simplify:
\begin{equation}
    H_{\rm min} = \ln{M} - \Delta \tilde{G} \left(1+\frac{e^{-\Delta \tilde{G}}}{1-e^{-\Delta \tilde{G}}}\right) + \ln\left(\frac{\Delta \tilde{G}}{1+e^{-\Delta \tilde{G}}}\right)+ 1. 
    \label{eq:Hcontinuous}
\end{equation}

\section{Minimizing entropy maximizes channel capacity}
\label{App:ChannelCapacity}
We can consider the deterministic CRNs in this paper to be information channels. Let us assume the underlying chemistry, which determines the rate constants appearing in the un-linearized model, and the concentration of monomers, are fixed. The effective rates of the linearized network would then vary with the concentrations of catalysts only. As a specific (simple) example of how the entropy bound defines the channel capacity, let us assume that all variability is due to $M$ sequence-specific template catalysts, one for each product, and that of these templates exactly one is present at any given time at a fixed concentration.

We can define the input to the information channel as the template that is present at high concentration; the output would then be a product sampled from the steady state product distribution for that input state. We can calculate the mutual information between the template input and output products. If the templates are symmetric, acting equivalently relative to their ideal sequence, then each output distribution would merely be a permutation of the set of product probabilities. 

A system of this kind would define a symmetric channel. The channel capacity of such a symmetric channel~\cite{Cover2006elements} is given by $C=\ln{M}-H([p_i])$ in our notation, where $M$ is the number of products/sequence-specific templates, and $H([p_i])$ the Shannon entropy of the output distribution for any single template. This entropy will be the same for any input state in our symmetrized description. Hence, minimizing the entropy maximizes the channel capacity.

\section{Example chemical reaction network that can saturate the bounds on accuracy}
\label{App:EX1}
Although calculations of the bounds on accuracy is often straightforward, evaluation of the actual performance of a system realisation can be more challenging. For the examples  in Sections~\ref{App:EX1} and~\ref{App:EX2}, steady state distributions are found by numerical solution of the underlying ordinary differential equations (ODEs). The CRNs are linear, connected and contain only one stoichiometric compatibility class. Hence, there exists a single positive steady state to the ODEs induced by mass action kinetics~\cite{polettini2014irreversible}. Initially, all concentrations are set to zero, and the ODEs are simulated for a large time until no change to the distribution is observed. For Section~\ref{App:EX2}, to speed up simulation, we make use a of a result from~\cite{qureshi2023universal} whereby we may coarse grain some sets of reactions without changing the steady state.

Refs~\cite{poulton2019nonequilibrium,qureshi2023universal} studied a system in which  a polymer is grown on a template. Monomers of either the ``right'' or the ``wrong'' type are added one at a time to a polymer in contact with a template, while the product polymer continually unbinds from the template from behind its leading edge. We now extend the system to include final dissociation from the template for complete polymers, and also a ``destructive template".  This destructive template participates in identical reactions to the template, except that polymerisation is driven backwards due to consumption of fuel molecules, meaning products tend to be destroyed rather than grown. 

\begin{figure}
    \centering
    \includegraphics[scale=0.35]{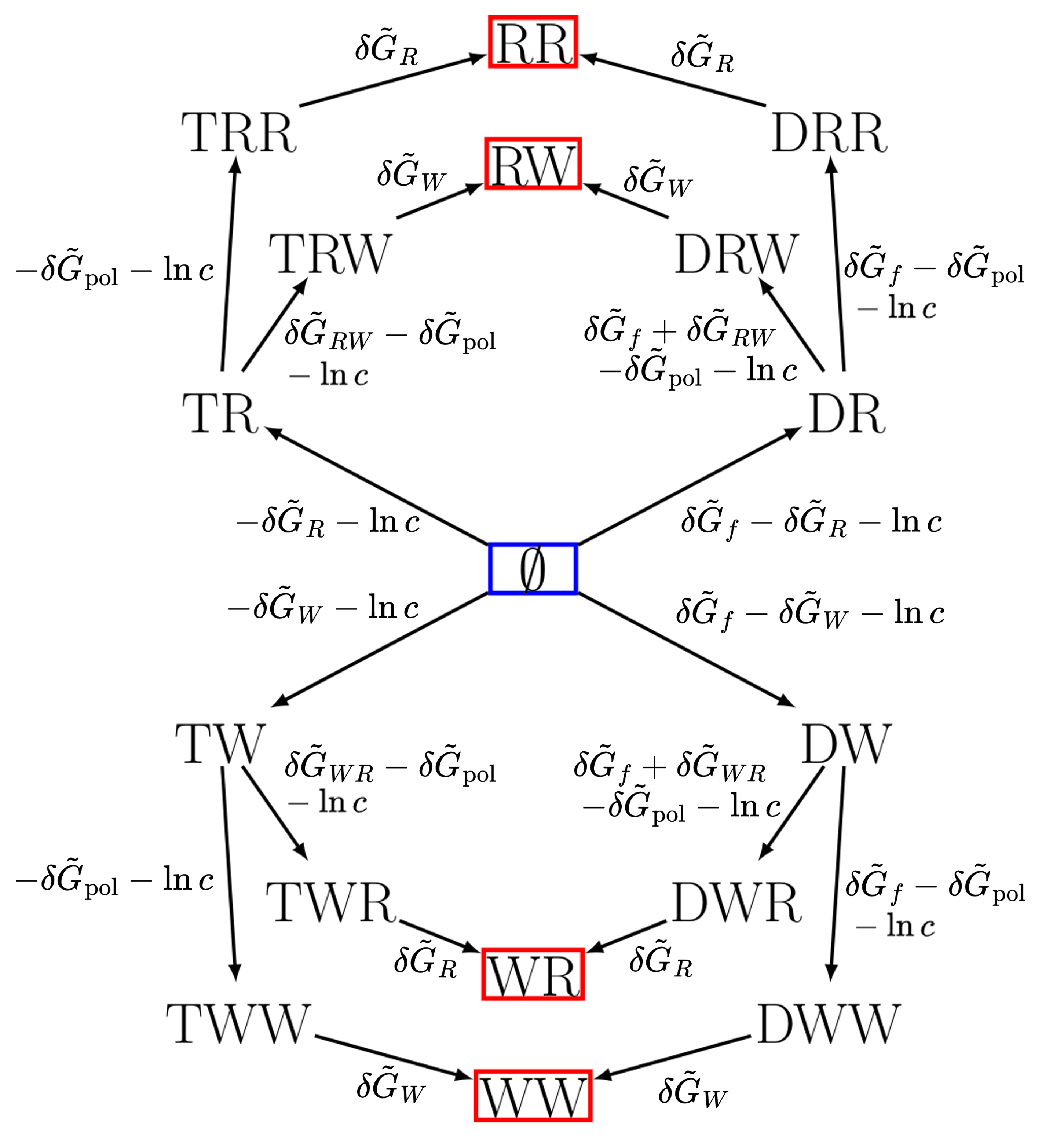}
    \caption{Linearised CRN for a system in which dimers are grown/destroyed via a template ``$T$" and by a destructive template ``$D$". The system starts in the null state (blue) and a right or wrong (``$R$" and ``$W$") monomer attach to either the template ``$T$" or destructive enzyme ``$D$". A second monomer
    can attach and polymerize with the first, yielding to a dimer that then detaches, giving the four red output states. Each arrow represents a reversible reaction, with the free-energy change in direction of the arrow indicated. For brevity, $\delta \tilde{G}_{RW}=-\delta \tilde{G}_{WR}=\delta \tilde{G}_R-\delta \tilde{G}_W$.  To reach each red product node, there are four self-avoiding walks from the blue null state.}
    \label{fig:DestroyerMetabolic}
\end{figure}

A full CRN for this model is shown below; we illustrate the linearised network (assuming monomers and catalysts are coupled to chemostats) for the case of dimerisation in figure \ref{fig:DestroyerMetabolic}. 
\begin{eqnarray}
    \nonumber T+X_1 &\xrightleftharpoons[k^T_{X_1}e^{-\delta \tilde{G}_{X_1}}]{k^T_{X_1}}& TX_1\\\nonumber
  TX_1\cdots X_{n-1}+X_n &\xrightleftharpoons[k^T_{X_1\cdots X_n}e^{-\delta \tilde{G}_{X_n}-\delta \tilde{G}_{\text{pol}}}]{k^T_{X_1\cdots X_n}e^{-\delta \tilde{G}_{X_{n-1}}}}&TX_1\cdots X_{n}\\\nonumber
  TX_1\cdots X_{L}&\xrightleftharpoons[k^{T,u}_{X_1\cdots X_L}]{k^{T,u}_{X_1\cdots X_L}e^{-\delta \tilde{G}_{X_L}}}&T+X_1\cdots X_{L}\\\nonumber
  D+X_1 &\xrightleftharpoons[k^D_{X_1}e^{\delta \tilde{G}_f-\delta \tilde{G}_{X_1}}]{k^D_{X_1}}& DX_1\\\nonumber
  DX_1\cdots X_{n-1}+X_n &\xrightleftharpoons[k^D_{X_1\cdots X_n}e^{\delta \tilde{G}_f-\delta \tilde{G}_{X_n}-\delta \tilde{G}_{\text{pol}}}]{k^D_{X_1\cdots X_n}e^{-\delta \tilde{G}_{X_{n-1}}}}&DX_1\cdots X_{n}\\\nonumber
  DX_1\cdots X_{L}&\xrightleftharpoons[k^{D,u}_{X_1\cdots X_L}]{k^{D,u}_{X_1\cdots X_L}e^{-\delta \tilde{G}_{X_L}}}&D+X_1\cdots X_{L},\\
  \label{eq:DestroyerCRN}
\end{eqnarray}
for $n\leq L$, $X_i\in\{R,W\}$. Dynamics is assumed to follow mass action kinetics, with rate constants given above and below the harpoons. $T$ represents the template, $D$ the destructive catalyst and $R,W$ the ``right" and ``wrong" monomers. The products are $X_1\cdots X_L$, representing the polymers of length $L$. Here, there are $M=2^L$ different products. $TX_1\cdots X_{n}$ and $DX_1\cdots X_{n}$ represent partial polymers, $X_1\cdots X_{n}$, bound to the template or destructive catalyst. Monomer $X$ binds to the template or destructive catalyst with standard free-energy $-\delta \tilde{G}_X$ and the standard free-energy of polymerisation is $-\delta \tilde{G}_{\text{pol}}$ in the absence of fuel. The destructive catalyst has an additional free energy $\delta \tilde{G}_f$ per length driving the disassembly of polymers.

The thermodynamics of the model are characterised by the standard polymerisation free energy of the monomers ($-\delta \tilde{G}_{\rm pol}$); the standard free-energy change of binding to the template for right and wrong monomers ($-\delta \tilde{G}_R$ and $-\delta \tilde{G}_W$), and the free energy of fuel turnover $-\delta \tilde{G}_f$. We assume that both right and wrong monomers are held at concentration $c$.

For the model depicted in figure~\ref{fig:DestroyerMetabolic}, we can calculate the free-energy change for different paths to each product state. For example, the paths from the null state to $RR$ are:
\begin{eqnarray}
    \begin{tabular}{l l l l l}
   $1)\;\;\emptyset$ &$\to TR$ &$\to TRR$ &$\to RR$, &\\
     $2)\;\;\emptyset$ &$\to TR$ &$\to TRW$ &$\to RW$ &$\to DRW$\\
     & &$\to DR$ &$\to DRR$ &$\to RR$,\\ 
    $3)\;\; \emptyset$&$\to DR$ &$\to DRW$ &$\to RW$ &$\to TRW$\\
    & &$\to TR$& $\to TRR$ &$\to RR$,\\ 
    $4)\;\; \emptyset$ &$\to DR$ &$\to DRR$ &$\to RR$. 
\end{tabular}
\end{eqnarray}
These incur free-energy changes: 
\begin{eqnarray}
    1)\; &-(\delta \tilde{G}_{\text{pol}}&+2\ln{c}),\nonumber\\
    2)\; &-(\delta \tilde{G}_{\text{pol}}&+2\ln{c}),\nonumber\\
    3)\; &-(\delta \tilde{G}_{\text{pol}}&+2\ln{c}) +\delta \tilde{G}_f,\nonumber\\
    4)\;&-(\delta \tilde{G}_{\text{pol}}&+2\ln{c}) +2\delta \tilde{G}_f.
\end{eqnarray}
Each pathway contains the terms $-(\delta \tilde{G}_{\text{pol}}+2\ln{c})$, corresponding to the standard free-energy change of product formation without any fuel turnover. Since our bounds only rely on the differences in free energies between pathways, we may drop these contributions, and the fuel free energy $\delta \tilde{G}_f$ alone determines the bounds. For polymers of length $L$, the equivalent standard free energy of product formation is $(L-1)\delta \tilde{G}_{\text{pol}}+L\ln{c}$, and it too may be dropped for consideration of the bounds.

Our bounds are achieved when rates of the path with the most negative free-energy change are maximised for the desired product(s), and rates of the path with the most positive free-energy change are maximised for all other products. For the $L=2$ case, for example, we could therefore maximise the rate of path 1 for $RR$, and the equivalents of path 4 for other products.  

\begin{figure}
    \centering
    \includegraphics[scale = 0.55]{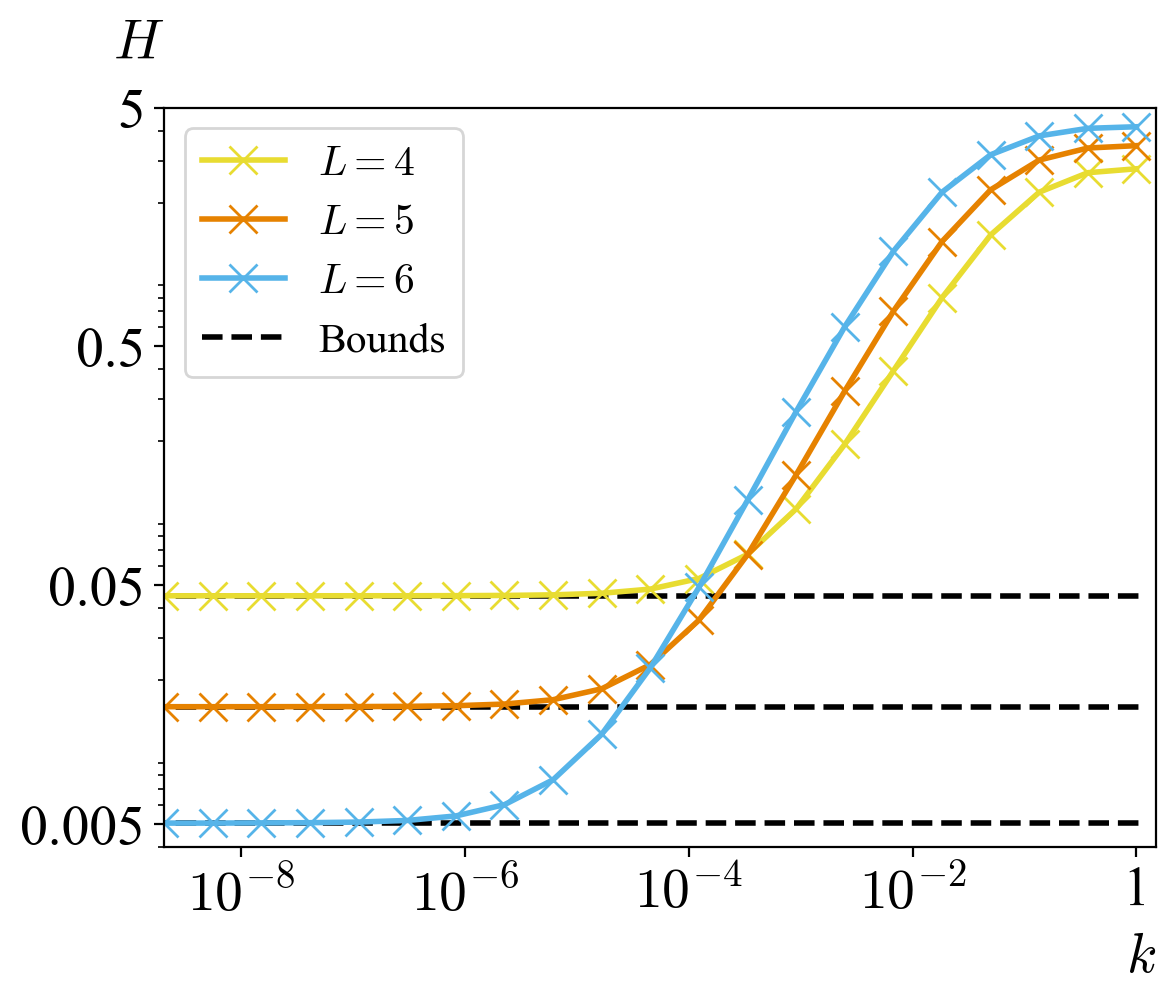}
    \caption{The entropy bound may be saturated by model systems in the limit that some reaction rates are much smaller than others. We plot the entropy, $H$, of the product distribution as a function of the slow reaction rates, $k$, for different template lengths, $L$, for the simple production and destruction model introduced in section~\ref{App:EX1}. The data is obtained for a fixed $\delta \tilde{G}_f=2$, $\delta \tilde{G}_{\rm pol}=0$, $\delta \tilde{G}_{R}=2$, $\delta \tilde{G}_W=-2$ and $c=1$. Note that longer templates reach a lower entropy bound for a fixed fuel turnover per unit length, $\delta \tilde{G}_f$.}
    \label{fig:SimpleHvsk}
\end{figure}

In fact, for this system, the entropy bound is formally achievable for arbitrary $L$. We can split the edges of the graph into two sets; one in which the rates are $\sim 1$, and one in which the rates are $\sim k$.
 In figure~\ref{fig:SimpleHvsk}, we show $H[p_i]$ as $k\rightarrow 0$ for a particular choice of these sets, in which all the reactions leading directly to the fully correct sequence being created/destroyed on the template are set to be fast (not proportional to $k$), as are all the reactions leading directly to the creation/destruction of other sequences on the destroyer.  Explicitly, set the rates of $\emptyset \to TR$, $TR \to TRR$, ..., $TR^{L-1} \to TR^L$,   $TR^L \to R^L$ equal to $1$ (here, $R^L$ corresponds to $L$ copies of $R$). We also set all the rates of $\emptyset \to DX_1$, $DX_1 \to DX_1X_2$, ..., $DX_1...X_{L-1} \to DX1...X_L$,   $DX_1...X_L \to X_1..X_L$, where $X_i=R$ or $W$ but excluding $X_1\cdots X_L$ all being $R$,  equal to 1. Conversely, we set the rates of $\emptyset \to DR$, $DR \to DRR$, ..., $DR^{L-1} \to DR^L$,   $DR^L \to R^L$ and $\emptyset \to TX_1$, $TX_1 \to TX_1X_2$, ..., $TX_1...X_{L-1} \to TX_1...X_L$,   $TX_1...X_L \to X_1..X_L$, where $X_i=R$ or $W$ but excluding $X_1\cdots X_L$ all being $R$, equal to $k_I=k$. The reverse reactions of those listed above have a rate determined by the free-energy change of reaction. For $k\to0$, this set of reaction rates saturates the bound. The system saturates the entropy bound $H_{\rm min}$ as $k\rightarrow 0$. Note that for the value of $\delta \tilde{G}_f$ used, the minimal entropy and maximal specificity distributions are the same. 

There are many possible ways to choose sets of edges that can saturate the bound in the limit $k\to0$. Here, we have chosen a set of rates that specifically highlights a full pathway to each product for illustrative purposes. One might also wish to choose a minimum set of reactions to have rate constant $k$ while still saturating the bound in the limit $k\to0$. For example, letting the slow reactions be $DR^L \to R^L$, where $R^L$ means $L$ copies of $R$, and $TX_1...X_L \to X_1..X_L$, excluding $X_1\cdots X_L$ all being $R$, will still saturate the bound in the limit $k\to0$. Further, we note that it is possible to saturate the bound with a non-specific destructive catalyst, where the reaction rates are independent of the polymer sequence. In the examples we have identified, such a network requires at least three rate scales $\sim 1,k,k^2$.

We stress that although the bound is formally attainable in this system, doing so relies on the ability to manipulate rate constants arbitrarily, subject to thermodynamic constraints. In a more realistic model of a templating system, constraints on relative rates may also be relevant; these constraints may stop the system reaching the bounds on accuracy or product entropy.

\section{Example chemical reaction network with kinetic proofreading that cannot saturate the bounds on accuracy}
\label{App:EX2}

To illustrate the application of the bound to a more complex network, and to demonstrate the possibility of non-trivial pathways defining the bound, we consider an extension to the previous model, wherein the template also performs kinetic proofreading. First suggested by Hopfield~\cite{hopfield1974kinetic} and Ninio~\cite{ninio1975kinetic} and widely studied \cite{bennett1979dissipation,mallory2020we}, kinetic proofreading is a mechanism by which a system can increase the specificity of a process by expending extra free energy through fuel consuming cycles. These cycles give an extra opportunity to reject the ``wrong'' monomers due to their shorter binding lifetime.

 The full chemical reaction network for a proofreading template of arbitrary length $L$ is:
 \begin{eqnarray}
     \nonumber T+X_1^* &\xrightleftharpoons[k^{T,I}_{X_1}e^{-\delta \tilde{G}_{X_1}}]{k^{T,I}_{X_1}}& TX_1^*\\
    \nonumber TX_1^* &\xrightleftharpoons[k^{T,\text{act}}_{X_1}]{k^{T,\text{act}}_{X_1}e^{\delta \tilde{G}_{\text{act}}}}& TX_1\\ 
     \nonumber T+X_1 &\xrightleftharpoons[k^T_{X_1}e^{-\delta \tilde{G}_{X_1}}]{k^T_{X_1}}& TX_1\\
     \nonumber TX_1\cdots X_{n-1}+X_n^* &\xrightleftharpoons[k^{T,I}_{X_1\cdots X_n}e^{-\delta \tilde{G}_{X_n}}]{k^{T,I}_{X_1\cdots X_n}}&TX_1\cdots X_{n-1}\circ X_{n}^*\\
     \nonumber TX_1\cdots X_{n-1}\circ X_n^* &\xrightleftharpoons[k^{T,\text{act}}_{X_1\cdots X_n}]{k^{T,\text{act}}_{X_1\cdots X_n}e^{\delta \tilde{G}_{\text{act}}}}&TX_1\cdots X_{n-1}\circ X_{n}\\
     \nonumber TX_1\cdots X_{n-1}+X_n &\xrightleftharpoons[k^T_{X_1\cdots X_n}e^{-\delta \tilde{G}_{X_n}}]{k^T_{X_1\cdots X_n}}&TX_1\cdots X_{n-1}\circ X_n\\
     \nonumber TX_1\cdots X_{n-1}\circ X_n &\xrightleftharpoons[k^{T,\text{pol}}_{X_1\cdots X_n}e^{-\delta \tilde{G}_{\text{pol}}}]{k^{T,\text{pol}}_{X_1\cdots X_n}e^{-\delta \tilde{G}_{X_{n-1}}}}&TX_1\cdots X_n\\
     \nonumber TX_1\cdots X_{L}&\xrightleftharpoons[k^{T,u}_{X_1\cdots X_L}]{k^{T,u}_{X_1\cdots X_L}e^{-\delta \tilde{G}_{X_L}}}&T+X_1\cdots X_{L}\\\nonumber
   D+X_1 &\xrightleftharpoons[k^D_{X_1}e^{\delta \tilde{G}_f-\delta \tilde{G}_{X_1}}]{k^D_{X_1}}& DX_1\\\nonumber
   DX_1\cdots X_{n-1}+X_n &\xrightleftharpoons[k^D_{X_1\cdots X_n}e^{\delta \tilde{G}_f-\delta \tilde{G}_{X_n}-\delta \tilde{G}_{\text{pol}}}]{k^D_{X_1\cdots X_n}e^{-\delta \tilde{G}_{X_{n-1}}}}&DX_1\cdots X_{n}\\\nonumber
   DX_1\cdots X_{L}&\xrightleftharpoons[k^{D,u}_{X_1\cdots X_L}]{k^{D,u}_{X_1\cdots X_L}e^{-\delta \tilde{G}_{X_L}}}&D+X_1\cdots X_{L},\\
   \label{eq:KPCRN}
 \end{eqnarray}

where $n\leq L$,\;$X_i\in\{R,W\},\;X^*_i\in\{R^*,W^*\}$. As in the previous CRN, $T$ represents the template, $D$ the destructive catalyst, $R$ the ``right" monomers and $W$ the ``wrong" monomers. $R^*$ and $W^*$ are non-activated monomers. Dynamics is assumed to follow mass action kinetics, with rate constants given above and below the harpoons. The species, $TX_1\cdots X_{n-1}\circ X_n$ ($TX_1\cdots X_{n-1}\circ X_n^*$) represent a complex of polymer of length $n-1$ bound to the template as well as a \mbox{(non-)activated} monomer $X_n$ ($X_n^*$) bound, but not yet polymerised into a single polymer of length $n$. $X_1\cdots X_L$ are the products. Both non-activated ($X^*$) and activated monomers ($X$) bind to the template or destructive catalyst with standard free energy $-\delta \tilde{G}_X$. If a non-activated monomer is bound to the template, it may be activated, with a free-energy change $\delta \tilde{G}_{\text{act}}$. If an activated monomer is bound to the template, it may be polymerised into the growing copolymer; the standard free-energy change of polymerisation is $-\delta \tilde{G}_{\text{pol}}$. The destructive catalyst has an additional free energy $\delta \tilde{G}_f$ per length driving the disassembly of polymers.

\begin{figure}
    \centering
    \includegraphics[scale=0.5]{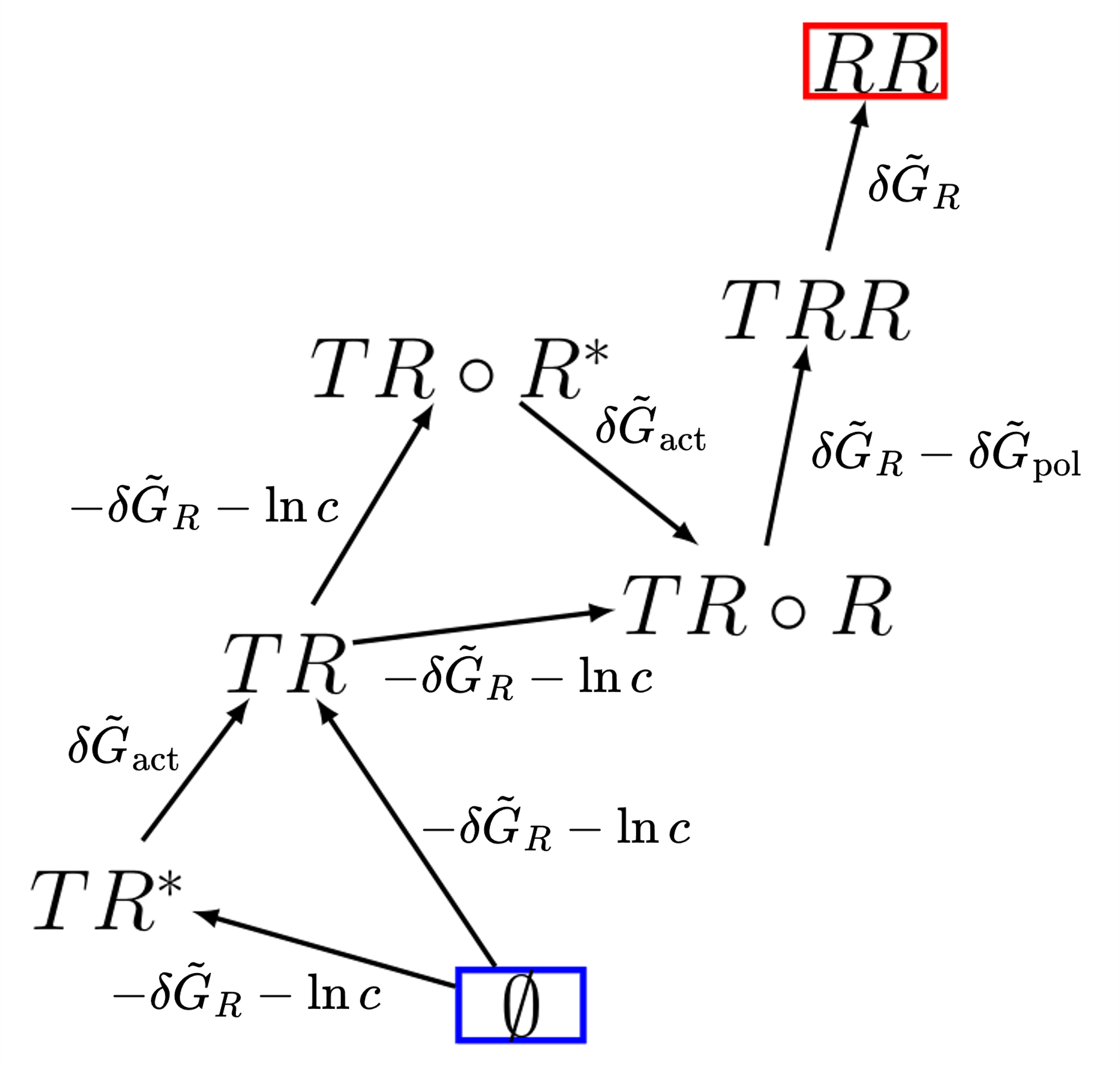}
    \caption{Modifying the template-based reactions of the model in section~\ref{App:EX1} to include  kinetic proofreading. We show here the template-based reactions leading directly to the dimer $RR$, which should replace the equivalent template-based reactions in figure~\ref{fig:DestroyerMetabolic}. As before, arrows represent reversible reactions with free-energy change in the direction of the arrow indicated. On the template, from state $TR$, either a non-activated ($R^*$) or activated ($R$) monomer may bind to the template. If that monomer has bound, but has not yet been polymerised into the growing polymer, it is represented by $TR\circ R^*$ or $TR\circ R$. When bound to the template, non-activated monomers may be activated, as shown by the transitions in which $R^*$ is converted to $R$. When there is an activated monomer at the end of the growing polymer ($TR\circ R$), that monomer may be polymerised into the growing polymer to reach a polymerised state ($TRR$). After a full length polymer has grown on the template, it may detach to a product ($RR$ for a dimer template).} 
    \label{fig:KPMetabolic}
\end{figure}

Once again, we linearise the system by assuming that monomers and catalysts are coupled to chemostats. In figure ~\ref{fig:KPMetabolic}, we show part of the linearised CRN graph; this fragment should be inserted into figure~\ref{fig:DestroyerMetabolic} in place of the pathway ($\emptyset\to TR\to TRR\to RR$), with similar modifications to all other template-based pathways to $RW$, $WR$, and $WW$. Proofreading adds complexity to the graph in the form of additional loops, and we now explicitly consider a polymerization step independently from the binding to the template.

Monomers are now present in inactive (starred) and active forms, with $-\delta \tilde{G}_{\text{act}}$, representing the free-energy change of activation. We assume that each non-activated monomer $R^*,\;W^*$ is chemostatted at the same  concentration as each activated monomer $R,\;W$, $c$. Dropping this assumption would only cause a shift to $\delta \tilde{G}_{\text{act}}$. As a result, $\delta \tilde{G}_f$ and $\delta \tilde{G}_{\text{act}}$ control the concentration bounds. We assume $\delta \tilde{G}_f,\;\delta \tilde{G}_{\text{act}}\geq 0$. 

\begin{figure*}[b!]
    \centering
    \begin{subfigure}{0.7\textwidth}
        \includegraphics[scale=0.5]{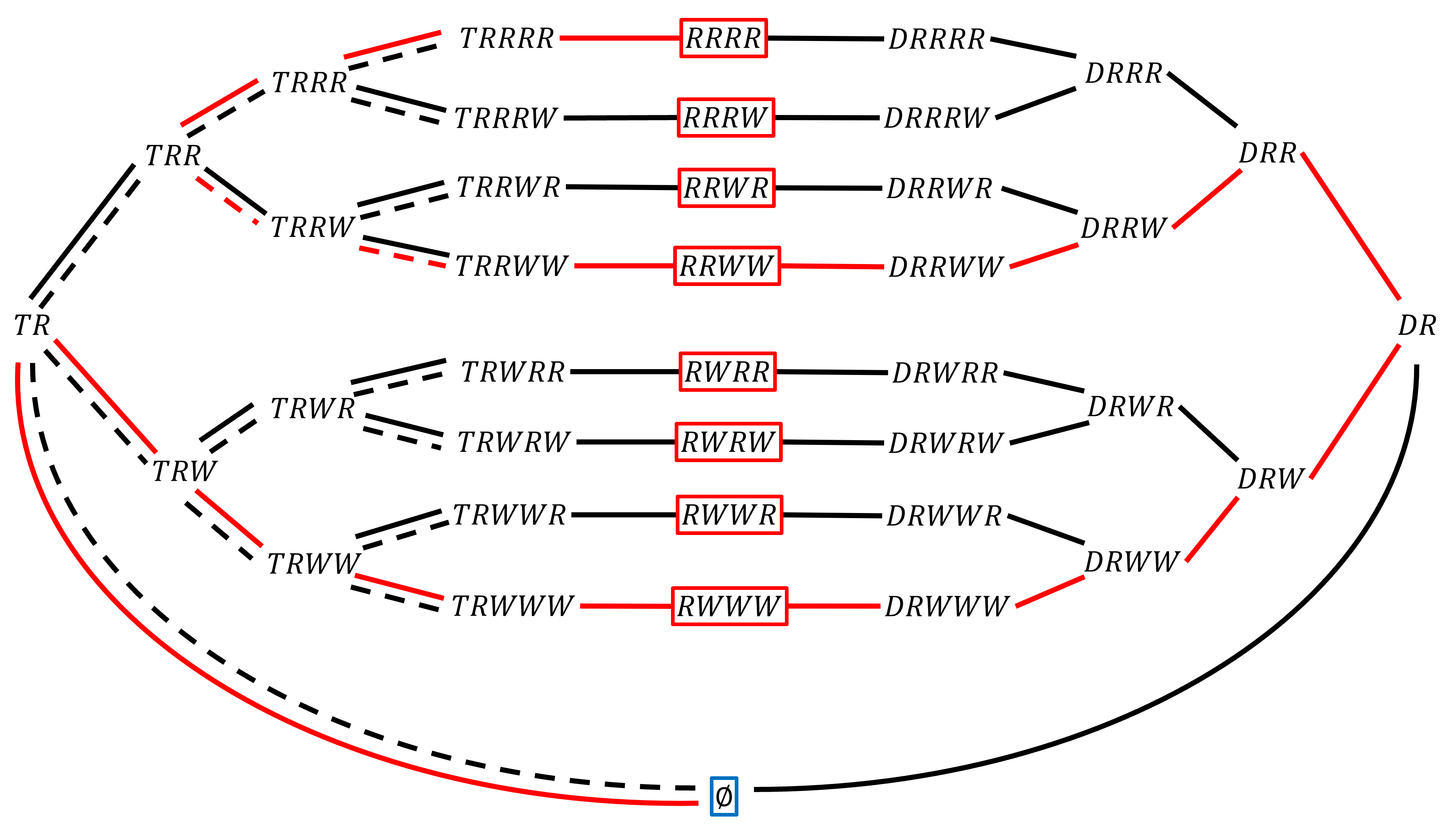}
        \caption{}
    \end{subfigure}\\
    \begin{subfigure}{0.7\textwidth}
        \includegraphics[scale=0.5]{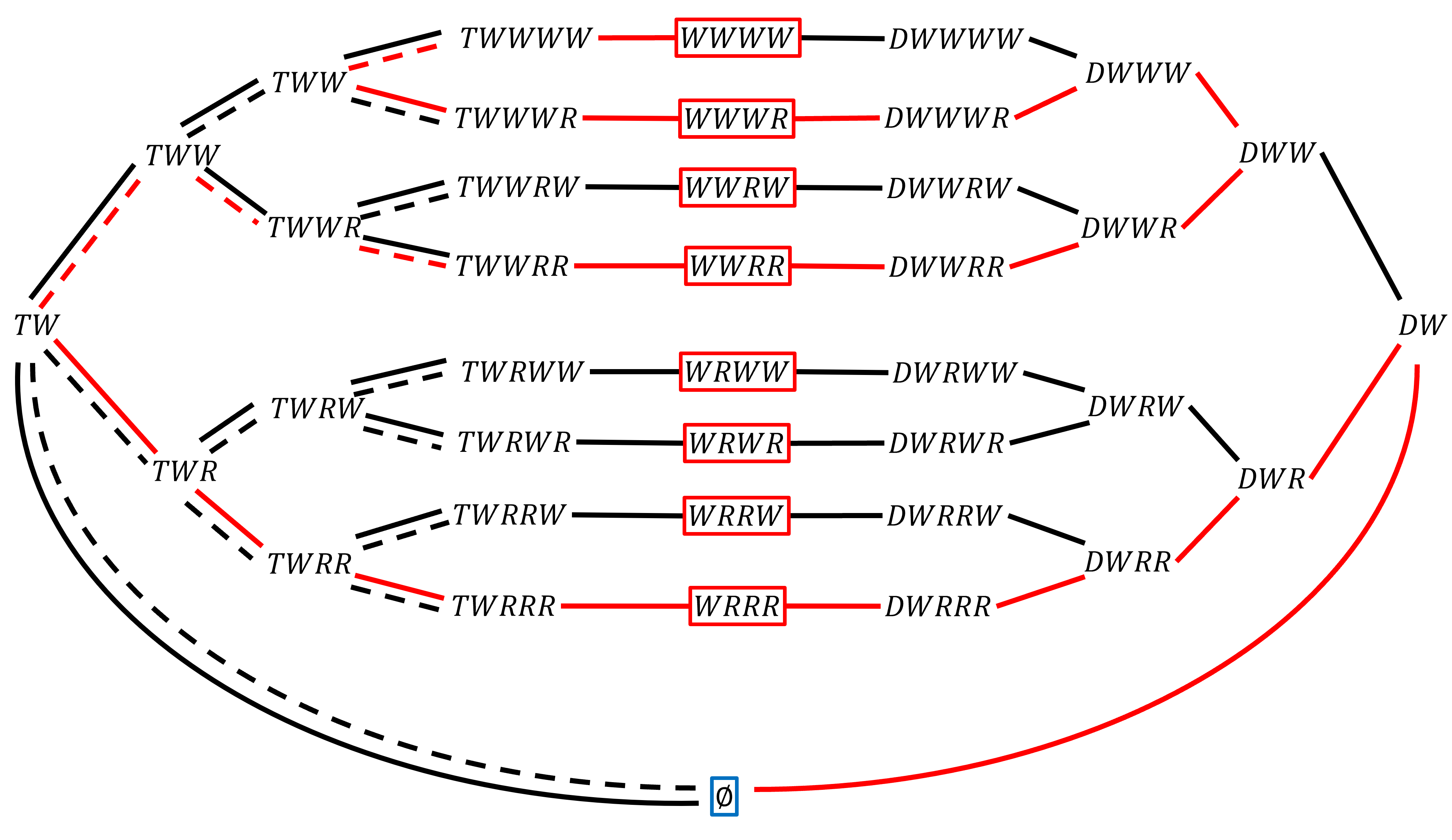}
        \caption{}
    \end{subfigure}
    \caption{(a) The pathway to $RRRR$ with free energy change $\delta \tilde{G}^*_U$ given by eq.~\ref{eq:KPUpperBound} is shown in red. (b) The pathway to $WWWW$ with free energy change $\delta \tilde{G}^*_L$ given by eq.~\ref{eq:KPLowerBound} is shown in red. For each of these diagrams, only the relevant half the reaction network is shown for simplicity. In these networks, there are two pathways between different template-bound states such as TRR and TRRR; one proceeding via $R^*$ and one bia directly binding to $R$ (Fig.~\ref{fig:KPMetabolic}). We represent these pathways via the solid and dashed lines, respectively.}
   \label{fig:KPUpLow}
\end{figure*}

The most extreme paths we have been able to identify for this system have free-energy changes
\begin{eqnarray}
    \delta \tilde{G}^*_U=\left\{
    \begin{array}{c}
          \frac{L^2}{4}+\frac{L}{2} \\
          \frac{(L+1)^2}{4}
    \end{array}
      \right\}\delta \tilde{G}_{\text{act}},
      \label{eq:KPUpperBound}
\end{eqnarray}
and
\begin{eqnarray}
        \delta \tilde{G}^*_L=-\left\{
    \begin{array}{c}
          \frac{L^2}{4} \\
          \frac{L^2-1}{4}
    \end{array}
      \right\}\delta \tilde{G}_{\text{act}}-L\delta \tilde{G}_f,
        \label{eq:KPLowerBound}
\end{eqnarray}
where the top value in each brace is for even $L$ and the bottom value for odd $L$. Unlike the simple system in section~\ref{App:EX1}, the SAWs that exhibit $\delta \tilde{G}^*_U$ and $\delta \tilde{G}^*_L$ are not the intuitively simple pathways that go via the template and the destroyer, respectively. Instead, the pathways correspond to snaking through the CRN, alternately using both the template and destructive catalyst to first create a sequence, and then convert it into another sequence. We show these pathways for $L = 4$ in  figure~\ref{fig:KPUpLow}.

Since we have not formally proved that $\delta \tilde{G}^*_L= \delta \tilde{G}_L$ and $\delta \tilde{G}^*_U=\delta \tilde{G}_U$, $\Delta G \geq \Delta G^*=\delta\tilde{G}^*_L-\delta\tilde{G}^*_U$ and the ``bounds'' implied by $\Delta \tilde{G^*}$ are not strict bounds on system performance. Rather, they are bounds on the bounds; if we could identify $\Delta G > \Delta G^*$, then our theory would allow for even better information propagation. Nonetheless, we have not identified any sets of parameters that allow even the level of accuracy implied by $\Delta G^*$ to be achieved. The SAWs yielding $\delta \tilde{G}^*_U$ and $\delta \tilde{G}^*_L$ would not only be absurd as the dominant pathways in a real system, they actually cannot dominate production and degradation, even in principle. Since the pathways exhibiting $\delta \tilde{G}^*_U$ and $\delta \tilde{G}^*_L$ pass through other products as intermediates, scaling these pathways to be fast would necessarily result in sub-pathways to other products being fast.

\begin{figure}
    \centering
    \includegraphics[scale = 0.55]{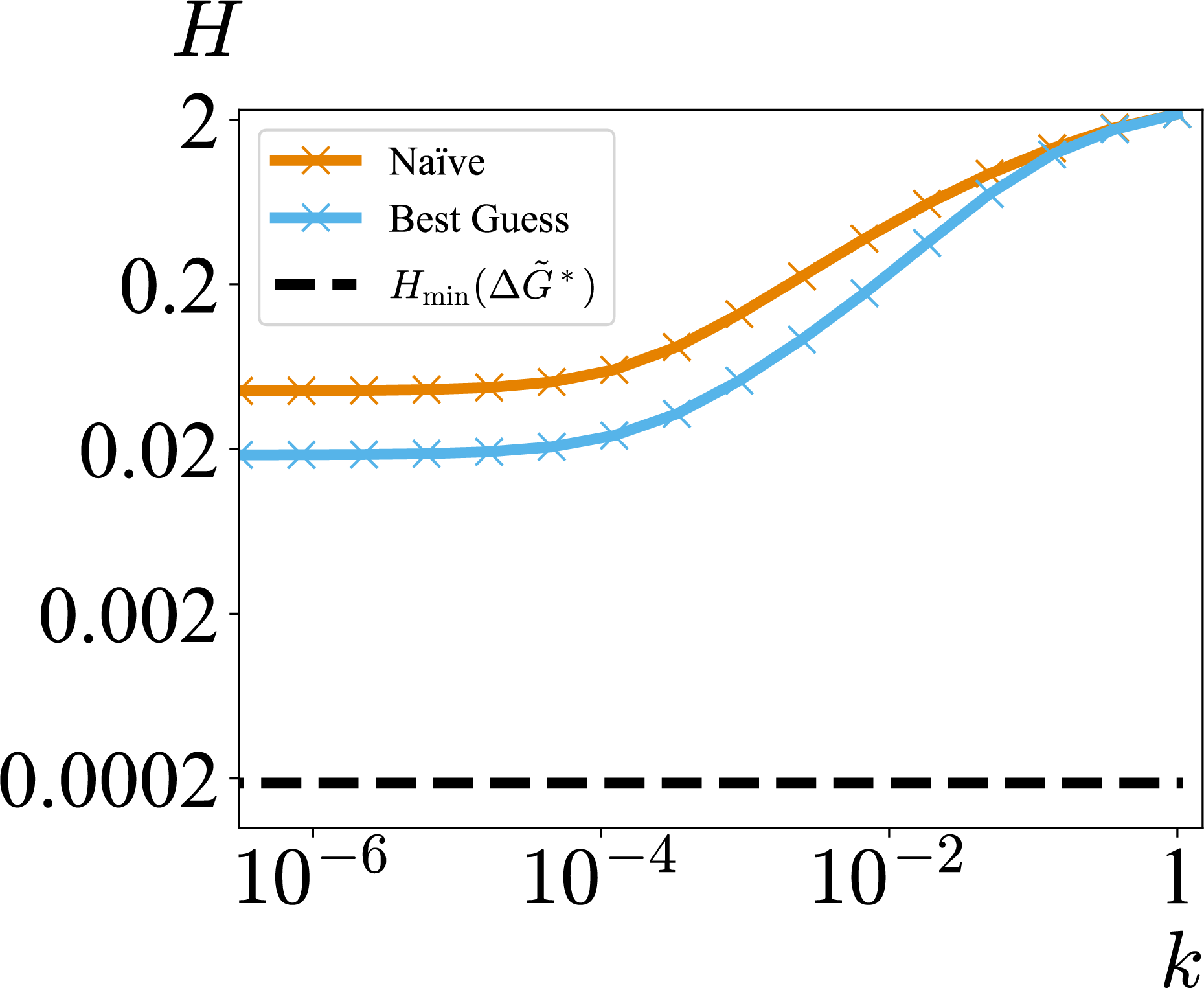}
    \caption{Entropy $H$ for two attempts to optimise the entropy, plotted alongside the lower bound $H_{\rm min}(\Delta \tilde{G}^*)$ for a system with $\Delta \tilde{G}=\Delta \tilde{G}^*$  implied by eqs.~\ref{eq:KPUpperBound}~and~\ref{eq:KPLowerBound}, as a function of the parameter $k$ that sets the overall scale of slow reactions relative to fast ones. In the ``na\"{i}ve'' approach, rates favour pathways to the correct product on the template and the incorrect ones on the destroyer. The ``best guess'' favours the snaking pathways described in text. The data is obtained for $\delta \tilde{G}_f=1$, $\delta \tilde{G}_{\rm act}=1$, $\delta \tilde{G}_{\rm pol}=0$, $\delta \tilde{G}_{R}=2$, $\delta \tilde{G}_W=-2$ and $c=1$.}
    \label{fig:KPHvsk}
\end{figure}

\begin{figure*}
    \centering
    \includegraphics[scale=0.5]{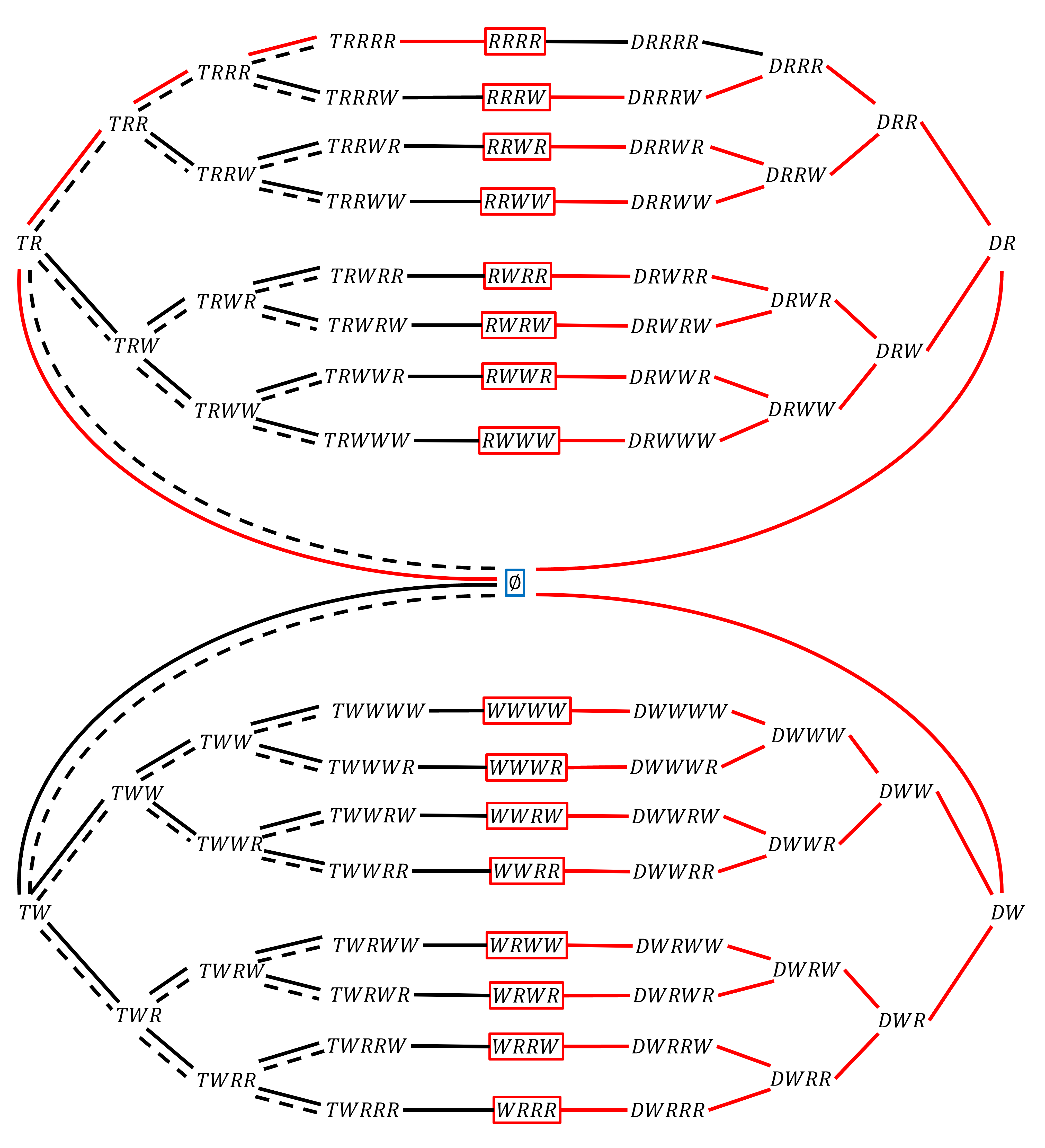}
    \caption{Fast reactions (red) with rate $\sim 1$ for the na\"ive attempt to minimize the entropy of the product distribution for a model with kinetic proofreading. In these networks, there are two pathways between different template-bound states such as TRR and TRRR; one proceeding via $R^*$ and one bia directly binding to $R$ (Fig.~\ref{fig:KPMetabolic}). We represent these pathways via the solid and dashed lines, respectively.}
    \label{fig:KPNav}
\end{figure*}

\begin{figure*}
    \centering
    \includegraphics[scale=0.5]{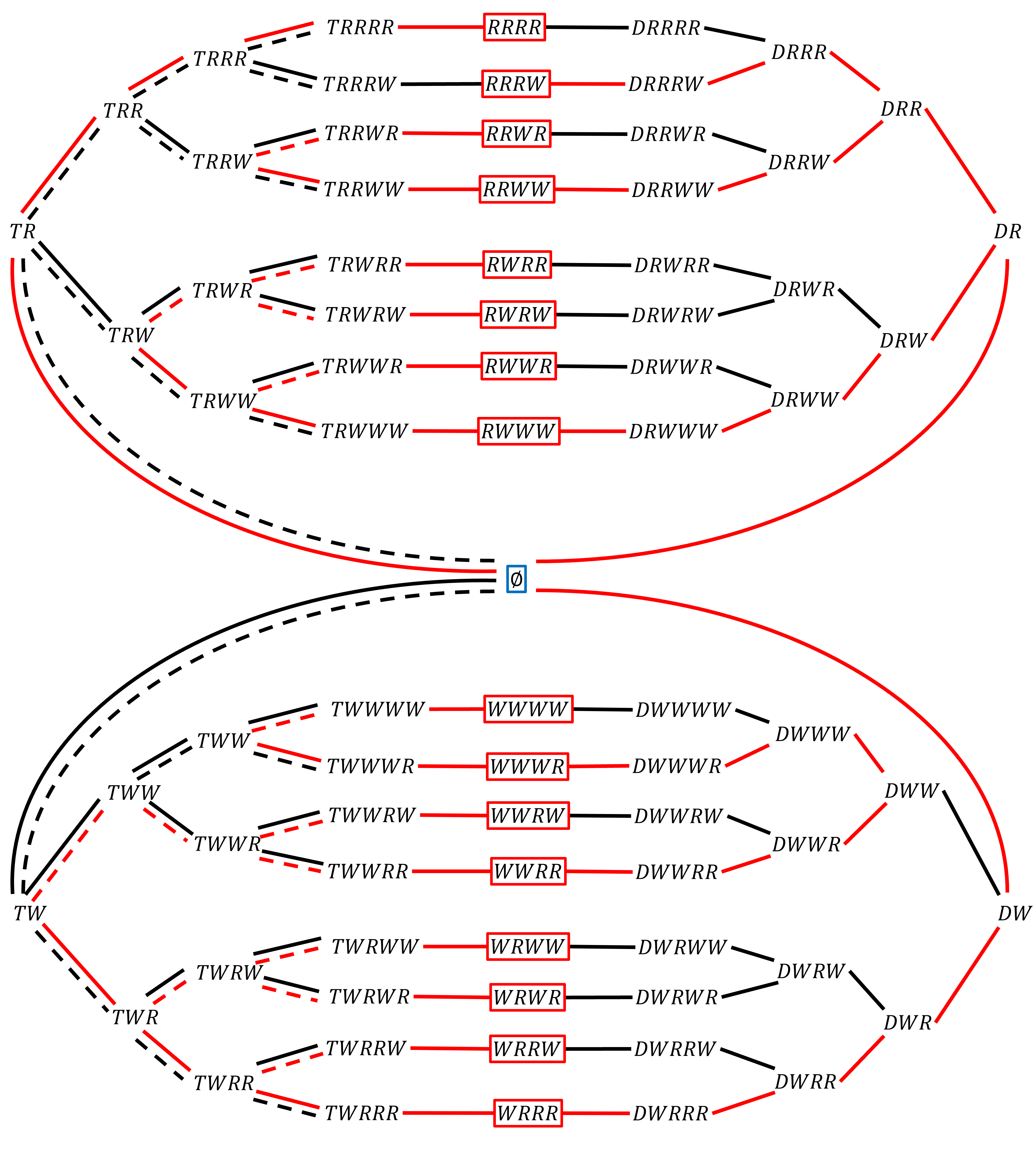}
    \caption{Fast reactions (red) with rate $\sim 1$ for our best guess at how to minimize the entropy of the product distribution for a model with kinetic proofreading. In these networks, there are two pathways between different template-bound states such as TRR and TRRR; one proceeding via $R^*$ and one bia directly binding to $R$ (Fig.~\ref{fig:KPMetabolic}). We represent these pathways via the solid and dashed lines, respectively.}
    \label{fig:KPBG}
\end{figure*}

Despite the above impossibility, the existence of these snaking pathways can provide some advantage, at least in principle. In figure~\ref{fig:KPHvsk}, we show two attempts to find parameters that minimize entropy for a system with $L=4$. In the first ``na\"ive'' scheme, plotted in blue, reactions contributing to assembly of $RRRR$ via the template or assembly of any other sequence via the destructive catalyst are assigned rates of $~1$ and all other rates are taken as $\sim k$. We show this scheme in figure~\ref{fig:KPNav}.
In the second ``best guess'' scheme, plotted in red, we make use of these snaking pathways. The reactions contributing to assembly of $RRRR$ via the template are still assigned rates of $~1$. However, for all other products, the longest snaking pathway which does not intersect with the pathway leading to $RRRR$ has rates assigned $~1$. We show this scheme in figure~\ref{fig:KPBG}.

As $k\rightarrow 0$, the snaking pathways outperform the non-snaking pathways significantly, resulting in approximately half the entropy of the product distribution (figure~\ref{fig:KPHvsk}). Notably, however, neither the best guess nor the na\"ive system converges on the minimal entropy of a system with a free-energy difference of  $\Delta \tilde{G}^*$ between pathways as $k \rightarrow 0$; the best guess approaches $H=1.8\times10^{-2}$\,nats as compared to the bound at $H_{\rm min}=1.9\times10^{-4}$\,nats.
Although the best guess system can outperform the na\"ive system in principle, when absolute rates can be chosen freely, rates may in practice be mechanistically constrained. As a result, this outperformance may not be achievable in a specific model. Indeed, at moderate $k$, the entropies of the naive system and the best guess converge.  



\end{appendix}

\bibliography{apssamp}

\end{document}